\documentclass[aps,prl,twocolumn,superscriptaddress,showpacs]{revtex4}

\usepackage{graphicx}

\begin{document}
\newcommand{\ket}[1]{\left|#1\right\rangle}


\title{Hydrogenic Spin Quantum Computing in Silicon: a Digital Approach}

\author{A. J. Skinner}
\email[]{Andrew.Skinner@Dartmouth.edu}
\affiliation{Laboratory for Physical Sciences, University of Maryland, College Park, MD, 20740}
\affiliation{Department of Physics and Astronomy, Dartmouth College, Hanover, NH, 03755}
\author{M. E. Davenport}
\affiliation{Department of Physics, California Institute of Technology, Pasadena, CA 91125}
\author{B. E. Kane}
\affiliation{Laboratory for Physical Sciences, University of Maryland, College Park, MD, 20740}


\date{\today}

\begin{abstract}
We suggest an architecture for quantum computing with spin-pair encoded qubits in silicon. Electron-nuclear spin-pairs are controlled by a dc magnetic field and electrode-switched on and off hyperfine interaction. This digital processing is insensitive to tuning errors and easy to model. Electron shuttling between donors enables multi-qubit logic. These hydrogenic spin qubits are transferable to nuclear spin-pairs, which have long coherence times, and electron spin-pairs, which are ideally suited for measurement and initialization. The architecture is scalable to highly parallel operation.
\end{abstract}


\maketitle

\section{}
A quantum computer comprising many two-level systems, or ``qubits,'' exhibits coherent superpositions (the incompatibility of certain observables) and entanglement (strong correlations between qubits).  These quantum features may be harnessed to solve problems which are essentially impossible for a classical computer, such as the factorization of large integers or the simulation of many-body quantum systems \cite{Nielsen}.  Solid state implementations stand to benefit from the rapid advances in semiconductor electronics and are potentially scalable to large arrays of qubits controlled by gate electrodes.  Donor nuclear spins in silicon are especially good solid state qubits because of their long coherence times.  They can in principle be controlled by hyperfine-tuned magnetic resonance techniques and coupled by the electron exchange interaction when carefully tuned surface gate voltages properly position the donors' electrons \cite{KaneQC}.  However, this ``exchange mediation'' is restricted to nearest neighbor interactions and is extremely difficult to control \cite{Burkard,Hu}; the coupling strength is very sensitive to the electrons' positions, exhibiting rapid oscillations due to Si band structure \cite{Andres,Koiller}.  Precise tuning of the hyperfine interaction will also be difficult.  In this paper we present an alternative donor spin architecture which tolerates tuning errors and overcomes nearest neighbor restrictions.

Our proposal relies on the ``encoding'' of each logical qubit, $\alpha\ket{0}+\beta\ket{1}$, in the $J_z=0$ subspace of a pair of spins: $\ket{0}\equiv(\ket{\uparrow\downarrow}-\ket{\downarrow\uparrow})/\sqrt{2}$ and $\ket{1}\equiv(\ket{\uparrow\downarrow}+\ket{\downarrow\uparrow})/\sqrt{2}$.  Encoding often results in reduced constraints on computer design \cite{LidarWu,DiVincenzo}.  When the two spins are donor nuclei the qubit benefits from their long coherence times.  On the other hand, measurements are facilitated when the two spins are electrons \cite{KaneSET,Recher}.  Following Levy, who proposed Heisenberg-only quantum computing with distinct magnetic moments in a static magnetic field \cite {Levy,LidarWu,Benjamin}, we will show that when the two spins are an electron and its donor nuclear spin (``a hydrogenic spin qubit'') the qubits are easier to control and can be coupled, well beyond their nearest neighbors, with electron shuttling.

\begin{figure}
\begin{center}
\includegraphics{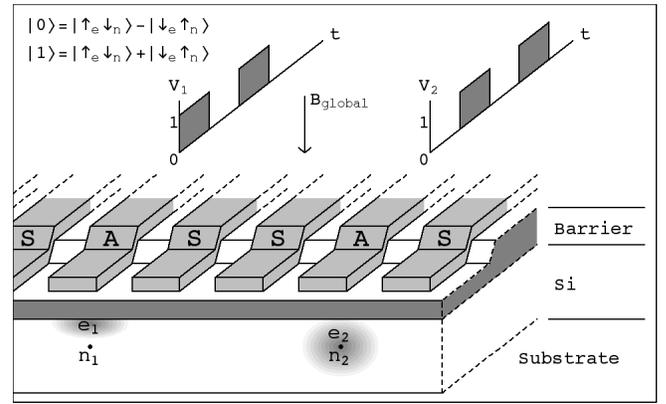}
\end{center}
\caption{Schematic of the proposed architecture.  Each qubit is encoded in the spins of an electron and its donor nucleus.  ``A-gates'' above donor sites switch the electron-donor overlap, and thus the hyperfine interaction, while ``S-gates'' shuttle electrons from donor to donor.  ``Bit trains'' of voltage pulses control the computer.}
\label{SchemArch}
\end{figure}

In the hydrogenic spin qubit the electron and donor nuclear spins are coupled by the hyperfine interaction.  The ground state coupling for P donors in Si, $H_A = A \,\vec{\sigma}_e\!\!\cdot\vec{\sigma}_n$, is ideally suited to quantum computing because it is a quadratic, and thus \it insensitive\rm, function of small external electric fields (perturbations from zero field); its strength, $A=121.517 \pm 0.021$ neV \cite{Feher}, is determined by the electron-donor overlap, $|\psi(0)|^2$.  Here $\vec{\sigma}\equiv(\sigma^x,\sigma^y,\sigma^z)$ are the Pauli operators, labeled by the spin on which they operate.  As depicted in Figure~\ref{SchemArch}, we can use a surface ``A-gate'' voltage to draw the electron off the nucleus, effectively switching off the coupling ($H_A \rightarrow 0$) to a regime which is similarly insensitive to tuning errors.  We therefore propose a digital approach \cite{Friesen}, in which the interaction is only on or off for each clock cycle as determined by ``bit trains'' of voltage pulses.  The hyperfine control generates the electron-donor ($e\!\!-\!\!n$) spin swap $\ket{0}+\ket{1} \leftrightarrow \ket{0}-\ket{1}$ and we augment this with  a globally applied static magnetic field, which generates $\ket{0}\leftrightarrow\ket{1}$.  For $\cal O\rm$(1 mT) fields the two generators are of comparable strength and an alternating series of interactions implements single qubit logic in direct analogy with Euler's theorem for constructing an arbitrary rotation from a sequence of rotations about distinct axes \cite{Nielsen}.

\begin{figure}
\begin{center}
\includegraphics{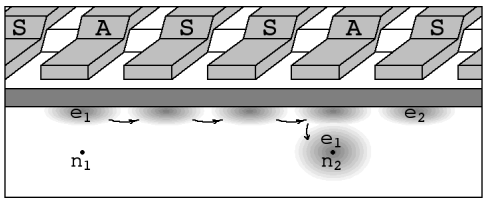}
\end{center}
\caption{Entangling qubits $e_1n_1$ and $e_2n_2$.  S-gates displace $e_2$ and shuttle $e_1$ to the vicinity of $n_2$.  The A-gate above $n_2$ then applies hyperfine interaction, generating a partial $e_1\!\!-\!\!n_2$ spin swap.}
\label{SchemEnt}
\end{figure}

Electron spin coherence distances of over $100\, \mu$m have been demonstrated \cite{Kikkawa}, so single electron shuttling \cite{Fujiwara} to remote donor sites is a good candidate for enabling two-qubit interaction.  As shown in Figure~\ref{SchemEnt}, arrays of ``S-gate'' electrodes between qubits are thus used to shuttle individual electrons from site to site.  Two qubits become entangled when the hyperfine interaction is applied between the electron of one qubit and the nucleus of another.  This is analogous to ion-trap proposals in which ions, and thus their quantum information, can be transported from one local trap to another \cite{IonTrap,Kielpinski}.  This transport is considerably more efficient than a bucket brigade series of nearest neighbor interactions and can circumvent misbehaved donor sites.

The evolution of the electron and donor spins is described by their Hamiltonian, \[
H = \sum_{i,j}A_{ij}\vec{\sigma}_{e_i}\!\!\cdot\vec{\sigma}_{n_j} + \sum_{i} B(g_e \mu_B \sigma_{e_i}^z - g_n \mu_N \sigma_{n_i}^z)
.\]  The second term, $H_B$, sums the contribution from all donors and their electrons, with respective magnetic moments $g_n \mu_N$ and $g_e \mu_B$, in the vertical magnetic field $B$ assumed parallel to a (100) lattice plane.  It augments the hyperfine contact term, $H_A$, which is a sum of interactions between electron-donor pairs.  Interaction between the $i$th electron and the $j$th donor is either off ($A_{ij}=0$) or on ($A_{ij}=A$).  We assume instantaneous switching and neglect the hydrogenic spin-orbit and dipole-dipole interactions (which are zero for the ground state and for sufficiently large $r$ but finite in between) as well as any randomness in the contact strength during the switch.  For P donors in Si the ground and first excited orbitals are separated by $\approx 15$ meV; a more realistic adiabatic switch takes $\cal O$(3 ps) which is fast compared to the hyperfine interaction.  Any remaining hydrogenic spin-orbit and dipole-dipole effects are coherent and can in principle be compensated by sophisticated control sequences or pulse shaping \cite{Cancel,PulseShape}, although we do not consider them here.  Similarly, we neglect the spin-orbit effect at the interface \cite{Rashba} because, for controlled shuttling of individual spins in Si, it is small, coherent, and, with further research, characterizable and correctable.

\begin{figure}
\begin{center}
\includegraphics{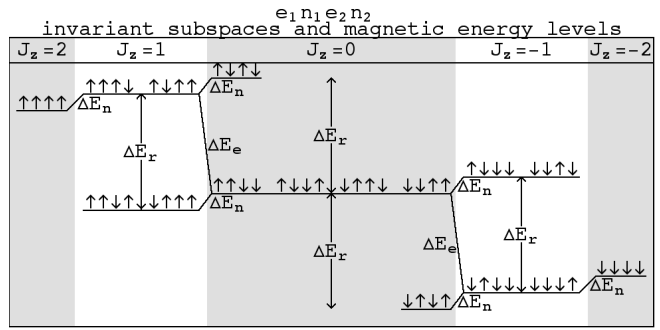}
\end{center}
\caption{Magnetic energy levels and invariant subspaces of a two-qubit computer.  Flipping a single electron or single nuclear spin changes the energy by $\Delta E_e$ or $\Delta E_n$ respectively and takes the state to another subspace.  Within an invariant subspace, simultaneous electron and donor spin flips change the energy by $\Delta E_r=\Delta E_e + \Delta E_n$.}
\label{invsubs}
\end{figure}

The state space of spins is decomposable into invariant subspaces labeled by the $z$ component of the total spin; up and down spins are stationary states of $H_B$ while electron-donor spin swaps, generated by $H_A$, preserve the number of up vs.\ down spins.  Within each invariant subspace flipping an electron spin, which changes the energy by $\Delta E_e = 2 B g_e \mu_B$ has a compensatory nuclear spin flip, which changes the energy by a further $\Delta E_n = 2 B g_n \mu_N$, and the magnetic energy splittings are thus integer multiples of $\Delta E_r = \Delta E_e + \Delta E_n$.  Transitions between subspaces require the flipping of one spin or the other and thus there exist nonresonant shifts $\Delta E_e$ and $\Delta E_n$ between subspaces.  As a specific example Figure~\ref{invsubs} shows the magnetic energy levels and invariant subspaces of a two-qubit computer.

It is desirable to generate pure hyperfine evolution even though the magnetic field is, in fact, always present.  We make use of the Trotter formula \cite{Nielsen},
\[
e^{-i H_A t/\hbar} \approx (e^{+i H_B \Delta t/2\hbar}e^{-i (H_A+H_B)\Delta t/\hbar}e^{+i H_B \Delta t/2\hbar})^a
,\] and compose a finite duration, $t$, of hyperfine evolution with a large number, $a$, of short $\Delta t=t/a$ steps of hyperfine \it and \rm magnetic evolution corrected, on the fly, by time-reversed $\Delta t/2$ steps of solely magnetic interaction.  Although magnetic and hyperfine steps do not commute, the remaining error of each step after correction, by a variant of the Campbell-Baker-Hausdorff formula \cite{Nielsen}, is $\cal O\rm(\Delta t^3)$ and increases with magnetic field (the non-commutivity, $[H_A,H_B]$, scales with $B$);  we can achieve good fidelity with sufficiently short $\Delta t$ steps and a sufficiently weak field.

Within each invariant subspace the time-reversed magnetic steps are achieved by incomplete periods of magnetic evolution.  A full period is determined by the energy splitting: $T_B=h/\Delta E_r$ (see Figure~\ref{invsubs}).  We need only wait for $T_B -\Delta t/2$ to achieve the magnetic correction step.  In analogy with magnetic resonance techniques we thus proceed by resonant stepping; for each period of magnetic evolution there is a short step of $H_A+H_B$.  The result is true hyperfine evolution up to relative phase shifts between invariant subspaces.

The use of digital bit trains from a pulse pattern generator considerably simplifies the timing of these operations.  For example, we divide the fixed hyperfine period, $T_A= h/4A = 8.50847$ ns, into $96$ clock cycles by setting the frequency at $f=11.2829$ GHz; given this frequency we then divide the magnetic period $T_B$ into $256$ clock cycles by choosing a field strength of $B=1.57171$ mT.  Within an invariant subspace, generating pure hyperfine evolution is now as simple as turning off certain A-gate voltages for 2 clock cycles out of every 256.

\begin{figure}
\begin{center}
\includegraphics{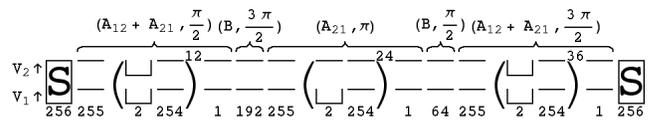}
\end{center}
\caption{Operations time line for an entangler.  Between shuttling operations represented schematically by an ``S,'' the diagram depicts a sequence of on-or-off A-gate voltages and their duration in clock cycles.}
\label{blank}
\end{figure}

The encoded qubits reside in the $J_z=0$ invariant subspace.  We can thus construct logic operations from finite $\phi$ pulses of magnetic evolution, $(B,\phi)\equiv e^{-i H_B \phi T_B/h}$, and $\theta$ pulses of pure hyperfine evolution, $(A,\theta)\equiv e^{-i H_A \theta T_A/h}$, implemented with resonant hyperfine stepping.  The Controlled-Not (CNOT) operation, which performs a logical NOT operation on a second qubit contingent on the state of a first, can, for example, be implemented in the following manner:
\[
CNOT=(L_1\otimes Z_2)N(L_1\otimes Z_2)^{\dagger},
\]
in which single qubit operations,
\[
(L_1\otimes Z_2) = (B,\frac{3\pi}{2})(A_{11}+A_{22},\pi)(A_{11},\frac{\pi}{2})(B,\frac{\pi}{2}),
\]
augment an entangler,
\[
N = (A_{12}+A_{21},\frac{3\pi}{2})
(B,\frac{\pi}{2})
(A_{21},\pi)
(B,\frac{3\pi}{2})
(A_{12}+A_{21},\frac{\pi}{2}).
\] This construction refines Levy's original \cite{Levy} but may not be optimal.  Figure~\ref{blank} depicts the actual sequence of A-gate voltages that implements the entangler, N.

Shorter hyperfine steps and a weaker magnetic field reduce the errors.  However, commercially available pulse pattern generators are limited to approximately 12 GHz (hence our choice of $f=11.2829$ GHz).  Furthermore, the preponderance of magnetic periods (one for each small hyperfine step) means that a computation slows with weaker field.  There is thus a trade-off between fidelity and speed.  Our choice of $B=1.57171$ mT yields a complete spin swap (the architecture's fundamental process) in $0.57 \mu$s.  
When ideally implemented with resonant hyperfine stepping, its expected error (defined to be the average probability of incorrectly transforming an initial, arbitrary, two-qubit basis of states) is less than $2.1\times10^{-7}$.  The CNOT is our most complicated gate and can be ideally implemented with an expected error of at most $0.9\times10^{-6}$ in $3.22 \, \mu$s.  

It is unrealistic to presume exact values for the frequency, field, and hyperfine strength.  There may also be variations of hyperfine and/or field strength from one donor site to the next.  Indeed, although isotope purification can remove most Si$^{29}$ from the crystal, the remaining impurities cause field variations (although these fluctuate so slowly that spin-echo techniques may be applicable).  Another complication is that the Land\'{e} factor for the electron, $g_e$, could vary by as much as $10^{-3}$ between the donor and the Si-barrier interface \cite{Feher}.

We have studied the sensitivity to these parameters by the explicit simulation of canonical one- and two-qubit logic gates.  The threshold theorem \cite{Nielsen} for quantum computation concludes that efficient quantum computing, obtained with error-correction techniques, is possible when logic gate errors are less than $10^{-5}$.  We found that this threshold is obtainable with relative variations in frequency, field, and hyperfine strength as large as $10^{-5}$, $10^{-5}$, and $5\times10^{-4}$, respectively.  The sensitivity to local variations in these parameters is approximately the same.  The fidelity is comparatively insensitive to the hyperfine strength because our gate compositions are predominantly magnetic.  Finally, the architecture can tolerate $5\times10^{-3}$ variations in $g_e$ between the donor and the interface.

A $\pi$ pulse of hyperfine interaction, $(A,\pi)$, between two qubits generates a complete spin swap between the electron of one qubit and the donor of the other.  Considered as a switch to a new encoding scheme, this hyperfine ``data bus'' transfers one qubit into a nuclear spin-pair and the other into an electron spin-pair.  For example, an $en$ data qubit, with the use of an $e_An_A$ ``ancilla,'' can be transferred, by resonant hyperfine stepping, into an $n_An$ nuclear spin-pair qubit.  Retrieval simply requires another $\pi$ pulse to repeat the spin swap.

The relatively weak nuclear magnetic moment gives the nuclear spin a long decoherence time which makes the nuclear spin-pair qubit a natural quantum memory.  Furthermore, if the data and ancilla were unentangled before the swap then the data (now encoded in the nuclear spin-pair) and ancilla (now encoded in the electron spin-pair) remain unentangled, so decoherence or collapse of the electron spin-pair will not degrade the memory (the qubit's transfer succeeds even when the ancilla is outside its logical subspace; relative phases developed between invariant subspaces, by resonant hyperfine stepping, are absorbed solely into the ancilla).

The data qubit can, alternatively, be transferred into an electron spin-pair to facilitate measurement by various proposed methods to distinguish singlets and triplets.  For an electron spin-pair known to reside in the logical subspace, these are effectively $\ket{0}=\ket{singlet}$ vs.\ $\ket{1}=\ket{triplet,S_z=0}$ projective qubit measurements.  For example, a Single Electron Transistor (SET) is capable of very sensitive charge configuration measurements; above a donor it can detect electrode driven charge density fluctuations associated with the electron spin-pair singlet \cite{KaneSET}.  Alternatively, in a quantum dot the electrons' spin determines the tunneling of spin-polarized currents \cite{Recher}.

\begin{figure}
\begin{center}
\includegraphics{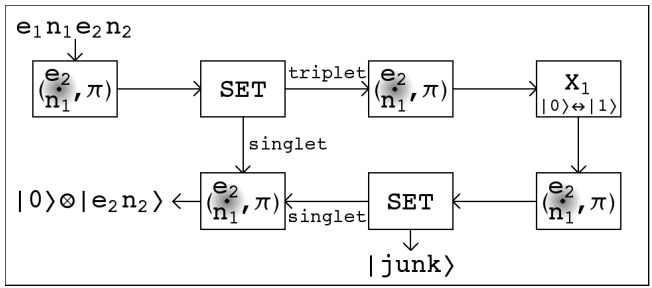}
\end{center}
\caption{Qubit Initialization and Sorting.  A singlet outcome is immediately convertible into $\ket{0}$ while the triplet outcome can be recycled through a sequence of operations into another chance for  a useful singlet.}
\label{Initial}
\end{figure}

After measurement the collapsed electron spin-pair can be transferred back into an electron-donor pair via another spin swap.  This provides a way to initialize the computer at high temperature (e.\ g., $1$ K).  Readout collapses an electron spin-pair into a singlet or triplet.  The singlet outcome, $\ket{\uparrow_{e_1}\downarrow_{e_2}}-\ket{\downarrow_{e_1}\uparrow_{e_2}}$, is immediately convertible, via a spin swap, to $\ket{0}$.  The triplet outcome, $\ket{\uparrow_{e_1}\uparrow_{e_2}}$, $\ket{\uparrow_{e_1}\downarrow_{e_2}}+\ket{\downarrow_{e_1}\uparrow_{e_2}}$, or $\ket{\downarrow_{e_1}\downarrow_{e_2}}$, can be recycled, as depicted in Figure~\ref{Initial}, through a single qubit $\ket{0}\leftrightarrow\ket{1}$ operation sandwiched between spin swaps, for another chance to obtain a useful singlet.  (This cascaded measurement prevails despite relative phases developed between invariant subspaces.)  At high temperature $50\%$ of the electron-donor pairs will obtain $\ket{0}$, and by electron shuttling the successful $50\%$ can be ``pooled'' into the working part of the computer in analogy with Kane's original proposal for on-chip spin refrigeration \cite{KaneRef}.

Hydrogenic spin qubits and coherent single electron shuttling enable a silicon-based quantum computer featuring digital hyperfine control insensitive to tuning errors, a long-lived nuclear spin memory, a projective readout scheme, and qubit refrigeration in which $50\%$ of the qubits can be initialized at high temperature.  The computer is scalable to highly parallel operation because digital shuttling of electrons overcomes nearest neighbor restrictions.  Finally, donors can be irregularly spaced and far apart, allowing for large gate electrodes, and malfunctioning donor sites can be diagnosed and avoided.  These many benefits motivate further research on the coherent shuttling and measurement of electron spins, extremely pure Si fabrication, encoding and error-correction techniques, optimal control sequences, and the spin-orbit and dipole-dipole interactions during realistic electrode driven switching and shuttling.

We are grateful for helpful discussions with S. Lomonaco.

\vspace{-1.125 in}
\subsection{}
\subsubsection{}

\bibliography{prl}

\end{document}